# A succession of relaxor ferroelectric transitions coexisting with ferroelectric states in $Ba_{0.55}Sr_{0.45}TiO_3$

Satyendra Singh, Satendra Pal Singh and Dhananjai Pandey[a]

School of Materials Science and Technology, Institute of Technology,

Banaras Hindu University, Varanasi-221 005, India

We present here the results of frequency dependent dielectric, polarization and powder X-ray powder diffraction studies in the 300 to 100K temperature range for $Ba_{0.55}Sr_{0.45}TiO_3$. The dielectric results indicate a succession of three relaxor ferroelectric transitions accompanying the cubic to tetragonal to orthorhombic to rhombohedral phase transitions confirmed by XRD studies. Our results confirm the coexistence of the relaxor ferroelectric and ferroelectric behaviours.

---

[a] Author to whom correspondence should be addressed;
electronic mail: dpandey@bhu.ac.in

Currently, there is enormous interest in relaxor ferroelectrics (RFE) not only from the point of view of the interesting physics involved but also from technological point of view for applications in multilayer capacitors and actuators.[1-3] It is believed that an understanding of the RFE state may provide a conceptual link between regular ferroelectrics on one hand and dipole glasses on the other. Most of the current studies on relaxors have been centered round $Pb(Mg_{1/3}Nb_{2/3})O_3$ (PMN) and other related compounds in which the average tetravalency of the 'B' site is maintained by the occupancy of lower and higher valent cations in an appropriate ratio. The observation of short range chemically ordered (SRCO) regions[4] in such compounds leads to quenched random electric field (RF) embedded in a disordered matrix.[5] The matrix shows a soft mode, above a characteristic temperature $T_D$, called Burn's temperature, which gets overdampded in the ergodic RFE state until it reappears as a hard mode in the non-ergodic RFE state below a freezing temperature, usually called Vogel-Fulcher freezing temperature ($T_{VF}$).[6] The question under intense debate at present is whether the non-ergodic RFE state in zero dc bias field is a random field induced nanodomain ferroelectric state[5,7] in the sense of the Imery and Ma's seminal work[8] or a dipolar glass state with randomly interacting polar nano regions in the presence of random fields[9].

Nearly a decade ago, work was initiated in our laboratory on a different route to RFE state in which isovalent substitutions above a critical concentration in regular FEs like $BaTiO_3$[10-12] and $PbTiO_3$[13] lead to frequency dependent permittivity maxima akin to archetypal RFEs like PMN. Unlike PMN and other similar compounds in which there is always local imbalance of ionic charges due to $Mg^{2+}$ and $Nb^{5+}$ occupancies at the B-site giving rise to quenched intense random fields, the isovalent substitutions do not lead to such a local charge imbalance. It is expected that the physics of such relaxor systems may not be identical to PMN type materials. In $Sr^{2+}$ substituted $BaTiO_3$, $Ba_{(1-x)}Sr_xTiO_3$ (BST), it was earlier shown that the permittivity maxima related with the cubic to tetragonal transition of $BaTiO_3$ shows relaxor features for $x \geq 0.12$ [10-12]. A similar relaxor behavior in BST has been reported at the $Sr^{2+}$ rich end also[14] in which experimental evidence exits for the presence of local polar regions in the paraelectric phase[15]. Relaxor behaviour has also been reported for $Ba(Ti_{1-x}Zr_x)O_3$ with $x \geq 0.27$ [16]. One of the outstanding issues in this class of relaxors based on $BaTiO_3$ is as to what happens to the succession of cubic (C) to tetragonal (T) to orthorhombic (O) to rhombohedral (R) phase transitions in these relaxor compositions. The purpose of this letter is to address this question in $Ba_{0.55}S_{0.45}TiO_3$ (BST45) using low temperature dielectric, polarization and XRD measurements. Our dielectric studies show that the permittivity maxima temperatures associated with the three transitions are frequency dependent akin to RFEs but also show clear signature of symmetry breaking in the XRD data providing direct evidence for coexistence of RFE and FE states conjectured earlier in thin films of BST[17] and postulated in several theoretical models.[18]

In the present work, chemically homogeneous BST45 powder was prepared by a semi-wet route described in ref. 10. XRD studies were carried out on a 18kW rotating anode (Cu- target) based Rigaku powder diffractometer (RINT 2000/ PC series) operating in the Bragg-Brentano geometry and fitted with a low temperature attachment and a curved crystal graphite monochromator in the diffracted beam. Structure of various phases was identified by Le-Bail technique using the FullProf[19] package. Dielectric measurements were carried out in the frequency range 100 Hz to 500 kHz at various temperatures (100-300 K) during cooling cycle using a LCR bridge (Hioki 3532 LCR Hitester).

The C-T, T-O and O-R transitions are clearly resolved in the temperature dependence of the real ($\varepsilon'$) and imaginary ($\varepsilon''$) parts of the dielectric constant of BST45 shown in Fig. 1(a). It is interesting to note that these transitions are quite diffuse. From a plot of $\ln(1/\varepsilon' - 1/\varepsilon'_m)$ versus $\ln(T - T'_m)$ in the paraelectric region shown in the inset to Fig. 1(a), one obtains an exponent of $\gamma = 1.8$ in agreement with similar values reported by previous workers[10,15] for various relaxor BST compositions. In addition, $T'_m$ and $T''_m$ shift to higher temperature side on increasing the measuring frequency as can be seen from Fig. 1(b-d) where the variation of $\varepsilon'$ and $\varepsilon''$ with temperature for the C-T, T-O and O-R transitions at a few selected frequencies are depicted. For pure BT, the transitions shown in Fig. 1 (a) are sharp with the mean field exponent $\gamma = 1$ and with $T'_m = T''_m$, where both $T'_m$ and $T''_m$ are frequency independent. The frequency dependent shifts of $T'_m$ and $T''_m$ along with the fact that $T''_m$ is always less than $T'_m$ for the C-T, T-O and O-R transitions in BST45 are akin to relaxor ferroelectrics[20] and dipole glasses.[21]

We have determined the temperature dependence of the relaxation time ($\tau$) for the relaxor type freezing associated with the C - T and O - R transitions from the positions of the peaks in $\varepsilon''$ versus T plots at various frequencies. No such analysis could be performed for the T - R transition, since the $\varepsilon''$ peak positions could not be located with sufficient confidence due to 'flat top' like nature of the broad profile, eventhough, qualitatively, the shifts of $T'_m$ and $T''_m$ to higher temperatures with increasing frequencies are quite evident. Fig. 2(a, b) shows the variation of $\ln\tau$ with temperature for Vogel-Fulcher type ($\tau = \tau_O \exp(\Delta E / k_B (T - T_{VF}))$) behaviour and the insets shows the corresponding Arrhenius type ($\tau = \tau_O \exp(\Delta E / k_B T)$) behaviour. Here, $\Delta E$ represents the barrier height (activation energy) for thermally activated jumps of polar clusters, while $k_B$ and $\tau_o$ are the Boltzman constant and inverse of the attempt frequency, respectively. It is evident from this figure that the relaxation process undergoes Vogel-Fulcher freezing for the both the C-T and O-R transitions, whereas the Arrhenius type freezing can be ruled out as the fit is not linear (see inset (i) in Fig. 2(a) and Fig. 2(b)). For Vogel-Fulcher behaviour, freezing temperature ($T_{VF}$) was taken as the temperature which gives minimum variance for the fit between $\ln(\tau)$ and (T- $T_{VF}$)[22]. This procedure gives $T_{VF}$, $\Delta E$ and $\tau_o$ as 235K, 3.07 x $10^{-3}$ eV and 5.6x$10^{-9}$ sec for the C-T transition and 128.5K, 3.4 x $10^{-3}$ eV and 6.0x$10^{-8}$ sec for the O-R transition. The values of $\Delta E$ and $\tau_o$ for the two transitions are comparable to those reported recently in another class of relaxors[23]. It is also evident that the plot of $\ln(\tau/\tau_o)$ vs (T-$T_{VF}$) shown in inset (ii) of Fig. 2(a) and Fig. 2(a) passes through the origin showing that the Vogel-Fulcher behaviour is due to critical freezing, expected for relaxors in the sense of ref. 24.

Fig. 3 shows the variation of remanent polarization ($P_r$) with temperature. It is evident from this figure that $P_r$ increases with increasing temperature in the rhombhohedral phase but in the orthorhombic phase, it is almost independent of temperature whereas after 184 K, in the tetragonal phase, it starts decreasing again. It is also clear from the figure that $P_r$ does not vanish at the T-C transition temperature but rather persists well above this temperature, as expected for relaxors[20] and in agreement with the earlier report of Singh et al[12] for x = 0.12.

Low temperature XRD studies confirm the existence of structural phase transitions in BST45 similar to those in pure BT. Figure 4 depicts the evolution of the 200, 220 and 222 pseudocubic reflections at 300, 210, 160 and 100 K. Full pattern Le-Bail refinements in the 2θ range 20-120° were carried out for the phase identification at these temperatures. The Le-Bail refinements unambiguously confirm the cubic, tetragonal, orthorhombic and rhombohedral structures at 300 K, 210 K, 160 K and 100 K, respectively, as can be inferred form the excellent fit between the calculated and observed profiles, shown in Fig. 4 with continuous line for a few selected reflections. Thus while the dielectric studies reveal relaxor type behaviour, the XRD results confirm the succession of C-T, T-O and O-R ferroelectric transitions suggesting clearly the coexistence of the relaxor ferroelectric and regular ferroelectric behaviours, as proposed in the thin film studies[17] but without any structural evidence.

Before, we conclude a brief comparison with PMN would be in order. In the archetypal relaxor ferroelectric PMN, the powder XRD patterns down to the lowest possible temperatures do not reveal any change in the average structure with respect to the cubic paraelectric state.[20] Further the difference between $T_{VF}$ and the main Curie peak at 1 kHz is ~ 40 K. However, in solid solutions of PMN with $PbTiO_3$ (PT), i.e. PMN-xPT, one observes not only clear signatures of symmetry breaking below $T_{VF}$, but also the difference $T_m^{/}$(10 kHz)–$T_{VF}$ gradually decreases and finally become zero for x = 0.35.[25] The compositions with 0.1≤ x < 0.35 exhibit on one hand frequency dependent $T_m^{/}$ and $T_m^{//}$ characteristic of the relaxor behaviour and on the otherhand clear signatures of symmetry breaking at powder diffraction level characteristic of a ferroelectric transition. This has been attributed to a gradual cross over from pure relaxor state for x < 0.10 to a pure ferroelectric state for x ≥ 0.35 with mixed features for the intermediate compositions[25]. The situation in BST45 appears to be similar to PMN–xPT with 0.1 ≤ x <

0.35 in the sense that there is a coexistence of both relaxor and ferroelectric features but with one very significant difference. Unlike PMN and PMN-xPT where there is only one frequency dependent dielectric peak, in BST45 one observes three dielectric peaks below room temperature, each one of them exhibiting relaxor behaviour. Evidently, the ergodic relaxor ferroelectric phase of BST45 first transforms to a non-ergodic phase below $T_{VF}$ with tetragonal symmetry. This non-ergodic phase 'reenters' into a ergodic relaxor state on further lowering of the temperature leading to another non-ergodic phase with orthorhombic symmetry, which again goes to a similar sequence on further lowering of the temperature leading finally to the non-ergodic ground state with rhombohedral symmetry. More work is needed to understand the origin of a succession 're-entrance'of relaxor ferroelectric transitions accompanied with ferroelectric ordering.

To summarise, we have shown that BST45 exhibits features common to both relaxor and ferroelectric states with evidence of a succession of relaxor transition associated with the cubic to tetragonal, tetragonal to orthorhombic and orthorhombic to rhombohedral phase transitions.

One of the authors (S P S) acknowledges financial support from the AICTE in the form of the award of a National Doctoral Fellowship.

**Figure captions:**

Fig. 1. Variation of the real ($\varepsilon'$) and imaginary ($\varepsilon''$) parts of the dielectric constant at (a) 10 kHz in the temperature range 100 to 290K and at the frequencies in the range 100Hz to 200kHz for (b) cubic to tetragonal transition (100 to 290K) (c) tetragonal to orthorhombic transition (150 to 230K) and (d) orthorhombic to rhombohedral transition (105 to 180K). The inset to Fig. 1(a) shows the variation of $\ln(1/\varepsilon' - 1/\varepsilon'_m)$ vs $\ln(T - T'_m)$ in the paraelectric region.

Fig. 2. Vogel-Fulcher fit for relaxation time corresponding to (a) cubic to tetragonal transition and (b) orthorhombic to rhombohedral transitions. Insets (i) to (a) and (b) shows the nonlinear nature of the $\ln\tau$ vs. $1/T$ plot, which clearly rule out Arrhenius behaviour. Insets (ii) to (a) and (b) show the variation of $1/\ln(\tau/\tau_o)$ vs $T-T_{VF}$.

Fig. 3. The temperature dependent variation of the remanent polarization ($P_r$).

Fig. 4. Observed (dots), calculated (continuous line), and difference (bottom line) profiles of the 200, 220 and 222 pseudocubic reflections obtained after Le-Bail refinements using x-ray powder diffraction data of BST45 in the $2\theta$ range 20 to 120°: (a) Cubic Pm3m space group at 300K, (b) Tetragonal P4mm space group at 210 K (c) Orthorhombic C2mm space group at 160 K and (d) Rhombohedral R3m space group at 100 K. The tick marks above the difference plot show the positions of the Bragg peaks for CuK$\alpha_1$ and K$\alpha_2$.

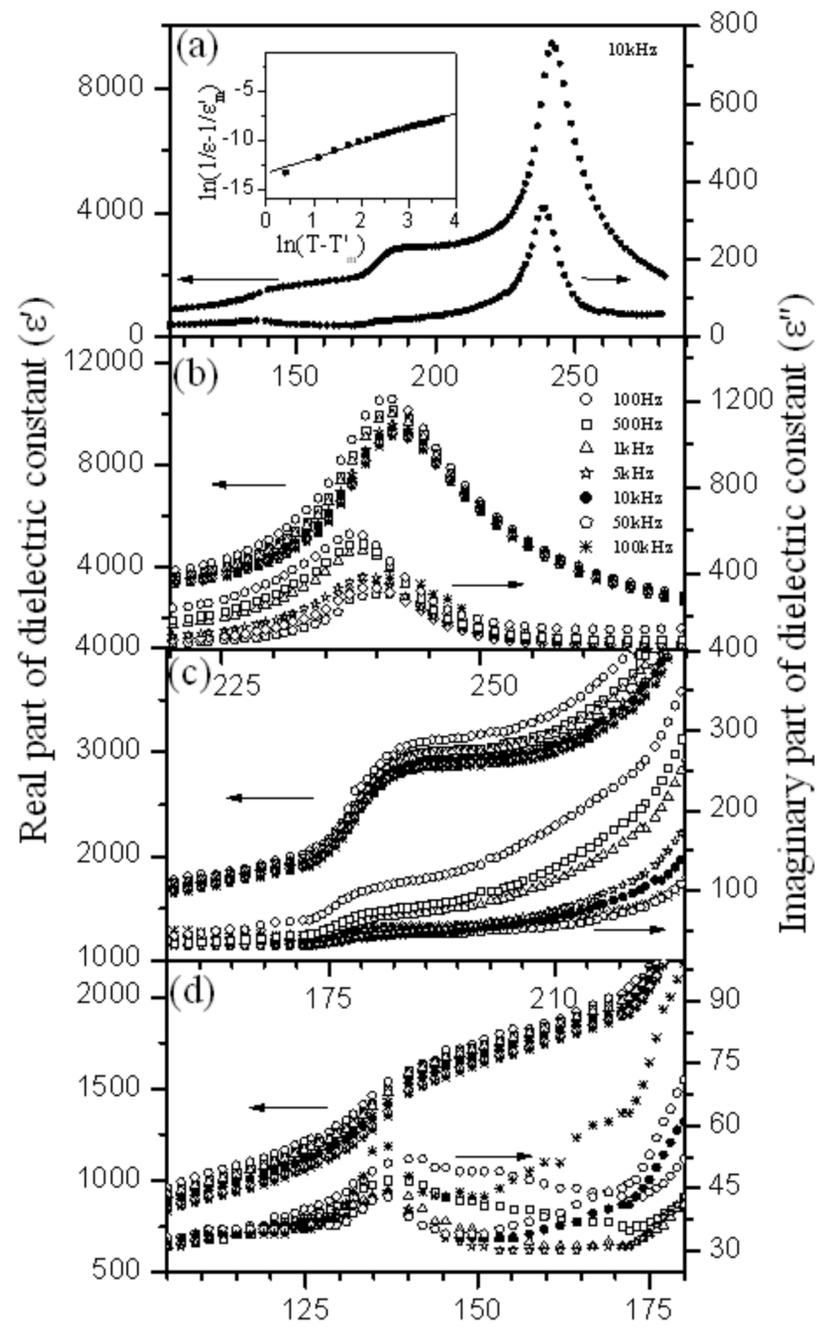

Fig. 1

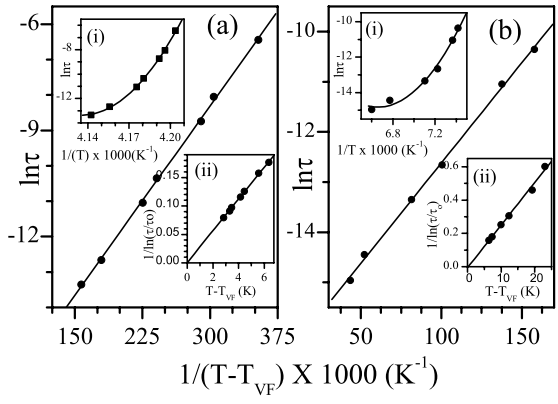

Fig. 2

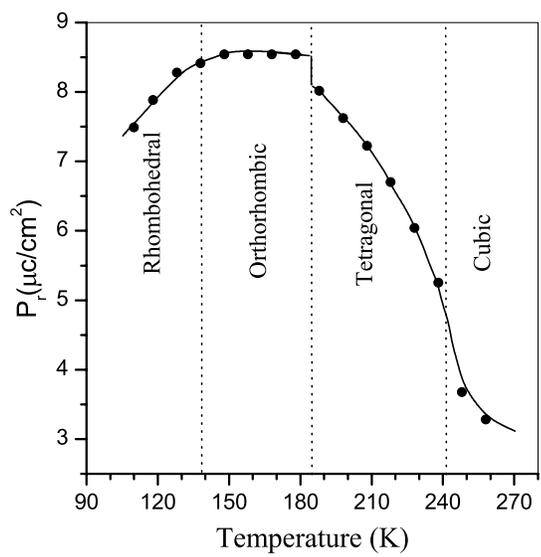

Fig. 3

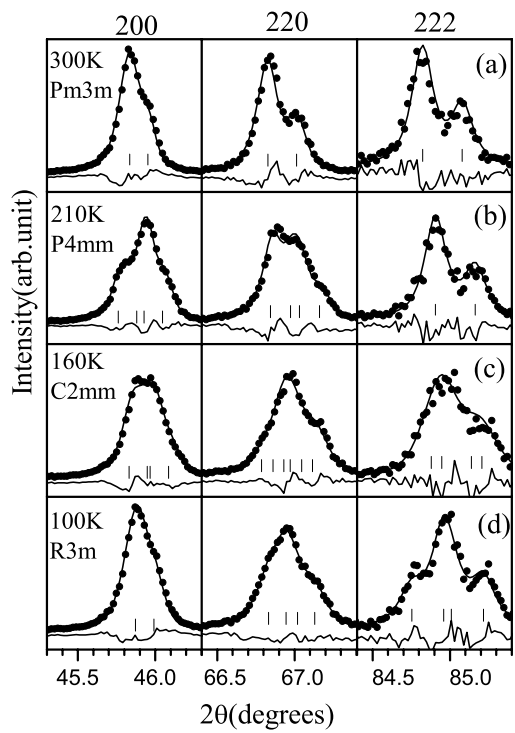

Fig. 4